\documentclass[twocolumn,showpacs,amsmath,amssymb]{revtex4}
                                                                                                  
\usepackage{graphicx,psfig}
\usepackage{dcolumn}
\usepackage{bm}
                                                                                                  
\begin{document}
                                                                                                  
\title{A Multi-level Algorithm for Quantum-impurity Models}
                                                                                                  
\author{Jaebeom Yoo}
\author{Shailesh Chandrasekharan}
\author{Harold U.\ Baranger}
\affiliation{Department of Physics, Duke University, Durham, NC 27708}

\date{\today}

\begin{abstract}
A continuous-time path integral Quantum Monte Carlo method using the 
directed-loop algorithm is developed to simulate the Anderson 
single-impurity model in the occupation number basis. Although the 
method suffers from a sign problem at low temperatures, the new 
algorithm has many advantages over conventional algorithms. 
For example, the model can be easily simulated in the Kondo limit
without time discretization errors. Further, many observables including 
the impurity susceptibility and a variety of fermionic observables 
can be calculated efficiently. Finally the new approach allows us
to explore a general technique, called the multi-level algorithm, to 
solve the sign problem. We find that the multi-level algorithm is able
to generate an exponentially large number of configurations with an 
effort that grows as a polynomial in inverse temperature such that 
configurations with a positive sign dominate over those with negative 
signs. Our algorithm can be easily generalized to other multi-impurity
problems.
\end{abstract}


\maketitle

\section{Introduction}

Developing efficient algorithms to solve problems in statistical 
mechanics involving strongly correlated fermions is an important 
area of research in computational physics. Such problems are often 
afflicted with sign problems and are difficult to handle using Monte 
Carlo techniques. The conventional approach is to integrate the fermions 
out and write the problem in terms of a statistical mechanics of a 
bosonic system where the fermionic physics is hidden in the Boltzmann 
weight as a determinant of a matrix \cite{Bla81,Hir}. In some interesting 
cases the 
determinant is positive definite which makes it possible to design
Monte Carlo methods for solving the problem. This approach is often 
referred to as the determinantal Monte Carlo method. Unfortunately, 
the method fails in many cases since the fermion determinant can often 
be negative. Even when the determinant is guaranteed to be positive 
the algorithms can be inefficient in certain interesting range of 
parameters. Determinantal Monte Carlo methods can often be formulated
only when the partition function is written as a path integral in 
discrete Euclidean time, which leads to time-discretization errors. 
Although such errors are controllable through extrapolation techniques 
they can be time consuming. All these difficulties make it important 
to explore alternative algorithms.

Recently, another approach to fermionic physics in which fermionic
partition functions are written in the occupation number basis has
gained popularity \cite{Wie93,Kaw94}. Although in this approach one 
encounters sign problems it is sometimes possible to design efficient 
algorithms in regions of parameter space where the sign problems are 
mild. In certain cases this approach also leads to novel solutions to 
the sign problems \cite{Cha99}, which in turn lead to algorithms which 
are much more efficient than conventional ones. This has lead recently,
for instance, to the
first successful confirmation of the Kosterlitz-Thouless behavior in
a fermionic model \cite{Cha02}. 
In this article, we explore a new algorithm to study the physics of
electrons in a partially filled band interacting with a few impurities.

Quantum impurity models are, of course, classics of condensed
matter many-body physics \cite{CMclassics}.  Interest in them has
been reawakened in recent years because of developments from two
completely different points of view.  On the one hand, quantum dots in
semiconductor heterostructures allow for the creation of tunable quantum
impurities which can be studied individually and with exquisite precision
\cite{qdots}.  Our particular interest is in studying the effects of
mesoscopic fluctuations on the many-body physics \cite{ribhu}.  On the
other hand, the study of strongly correlated electron systems away from
half filling -- systems which continue to be investigated intensively --
leads naturally to fermionic quantum impurity models through the dynamical
mean theory approximation \cite{dmft}.  We plan to use our method in
this connection in the future.  Since the Anderson single impurity
model is the simplest in this class \cite{And,Hew}, we focus on it;
it is straight forward to extend the method to include more impurities.

The Hamiltonian of the Anderson impurity model which we consider is
\begin{eqnarray}
H &=& 
\sum_{{\bf k},\sigma}\epsilon_{\bf k} c^\dagger_{{\bf k}\sigma} 
c_{{\bf k}\sigma}
  + \sum_\sigma \epsilon_dd^\dagger_\sigma d_\sigma \nonumber \\
  &+& \sum_{{\bf k},\sigma}V_{\bf k}(c^\dagger_{{\bf k}\sigma} d_\sigma
  + d^\dagger_\sigma c_{{\bf k}\sigma})
+ U d_\uparrow^\dagger d_\uparrow d^\dagger_\downarrow d_\downarrow, \label{aham}
\end{eqnarray}
where the first term represents the free band electron energy levels, 
the second term is the impurity energy, the third represents the hopping 
amplitudes between free electron states and the impurity, and the last 
term is the repulsive coulomb interaction term at the impurity site.
We assume that $-D \leq \epsilon_{\bf k} \leq D$ where $2D$ is the bandwidth.
This model was introduced over forty years ago by Anderson \cite{And} to 
study the affects of impurity spins embedded in metals. Today, this model
plays a very important role in understanding a variety of condensed matter
systems \cite{Hew}. The problem can
be solved analytically for the case of a constant hopping amplitude and 
a constant density of energy levels with an infinite bandwidth, using the
Bethe ansatz \cite{Wig,Kaw}. In the limit of large $U$ one can relate this 
model to the famous Kondo model \cite{Hew}. After his discovery of 
renormalization, Wilson used this model to illustrate the numerical 
renormalization group 
program \cite{Wil}, which is a powerful method to solve this problem.
It is now well understood that 
the low temperature properties of this model require a non-perturbative 
approach.

More than a decade ago Hirsch and Fye developed a determinantal Monte-Carlo 
algorithm to study this model \cite{Hir}. Such an approach is necessary to 
solve the Hamiltonian in Eq.(\ref{aham}) with general parameters. In this
method the partition function for a given temperature is rewritten as a
statistical mechanics of a model on a discrete temporal lattice. An 
auxiliary Ising variable for the impurity is introduced as a function of 
time in order to convert the Hamiltonian into a problem in which fermions 
are free but interact with the Ising spin. It is then possible to integrate 
the fermions completely to write the problem as a statistical mechanics of 
the Ising spin. The Boltzmann weight can be written in terms of certain 
Green functions which can be computed easily. Unfortunately, the 
algorithm slows down as the temporal lattice spacing is taken to zero for
a fixed temperature. It is also difficult to approach the large $U$ 
limit since the discretization error becomes significant. Finally, some
observables are difficult to evaluate. One famous example is the impurity 
susceptibility which is known to contain large fluctuations \cite{Fye}.

Here we explore an alternate approach, by writing the partition function 
in the occupation number basis in continuous Euclidean time. Since fermionic 
occupations consist of two states one can use the recently developed 
directed-loop algorithm for quantum spin systems \cite{Sand} in continuous
time \cite{Bea96}. Unlike the Hirsch and Fye algorithm, in our method one 
can only deal with a finite number of energy levels, but the discretization 
error in the Euclidean time direction can be eliminated. This allows us to 
simulate a large value of $U$ with little effort. In the occupation number 
basis we can also easily calculate observables such as the average occupation 
numbers, and the local susceptibility. Moreover, as we will discuss later 
we are able to calculate the impurity susceptibilities efficiently. 

Inspite of these advantages, the Boltzmann weights of our configurations 
can be negative due to the fermion permutation sign. Thus, our method 
suffers from a sign problem. However, we find that the sign problem is 
rather mild down to the Kondo temperature. Interestingly, our approach also 
allows us to explore a new technique, called the multi-level algorithm, 
to solve the sign problem. This technique was recently used in lattice QCD 
in determining the string tension between quarks and anti-quarks \cite{Lue}.
A similar technique was also explored in \cite{Mak98}. Since the multi-level 
technique is a general method, our method allows
us to study the usefulness of this approach to solve a class of fermion 
sign problems. Here we show that the multi-level technique is indeed useful in 
alleviating the sign problem. We were able to estimate signs of the order 
of $10^{-8}$ using this approach.

The paper is organized as follows: In section II we introduce the Monte Carlo
algorithm for updating the Anderson impurity model in the occupation 
number basis. In section III we explain the multi-level algorithm and show 
how one can use it to calculate the average signs efficiently. In section IV 
we discuss how one can measure observables in our method. We discuss the
calculation of the impurity susceptibility and how our method allows us
to compute it efficiently. Section V contains some results from the 
algorithm. Section VI contains our conclusions.

\section{The Directed Loop Algorithm}

In this section we construct a Monte Carlo method to calculate quantities
for the Anderson impurity model described by the Hamiltonian of 
Eq.(\ref{aham}). We begin by rewriting the partition function, 
$Z = {\rm Tr}\,e^{-\beta H}$ at temperature $T=1/\beta$, as a 
path integral in Euclidean time. This is accomplished by introducing
$M(\equiv\beta/\tau)$ imaginary time slices and writing
\begin{equation}
Z = {\rm Tr\,e}^{-\beta H} \approx \sum_{C}
        \prod^{M-1}_{i=0} <C_i|{\rm e}^{-\tau H}|C_{i+1}>, \label{eqn:part}
\end{equation}
where $C_i$ represent the electron states on the $i$-th time slice in 
the occupation
number basis. Since we are evaluating the trace we must have $C_0 = C_M$.
The true partition function in continuous time is obtained in 
the limit of large $M$ and small $\tau$ at fixed $\beta$. We can define
\begin{equation}
W[C] \sigma[C] \equiv \prod^{M-1}_{i=0} <C_i|{\rm e}^{-\tau H}|C_{i+1}>,
\label{boltz}
\end{equation}
where $W(C)$ is the magnitude and $\sigma[C]$ the sign of the
Boltzmann weight. Then the partition function can be written as
\begin{equation}
Z = \sum_{C} W[C]\sigma[C].
\end{equation}
In the Monte Carlo method each space-time configuration $C$ is generated 
stochastically with probability
\begin{equation}
P(C) = \frac{W[C]}{\sum_{C} W[C]}.
\label{prob}
\end{equation}
Ignoring the sign, each configuration of 
fermion occupation numbers is analogous to that of a configuration of a
quantum spin-half particles. Hence we can use an extension of the 
directed loop algorithm discussed in \cite{Sand} to update the space-time 
occupation number configurations $C$ directly in continuous time. A simple 
way to construct such an update is to construct the update rules for finite 
$M$ and then take the limit of infinite $M$. Below we describe our rules.

A special feature of the Anderson model is that electron hopping must
include the impurity and any of the band electron sites. However,
since all the band electron sites are involved, in our algorithm a 
``vertex'', in the language of \cite{Sand}, is change in the configuration between
two adjacent time slices, $C_i$ and $C_{i+1}$. If $C_i=C_{i+1}$ then one
has a diagonal configuration with weight
\begin{eqnarray*}
<C_i|{\rm e}^{-\tau H}|C_{i+1}> &=& 1- \tau \sum_{{\bf k},\sigma} \epsilon_{\bf k} n_{{\bf k}\sigma} \\ 
&-& \tau \sum_{\sigma} (\epsilon_d n_{d \sigma} + U n_{d\uparrow} n_{d \downarrow}),
\end{eqnarray*}
up to first order in $\tau$. Since we will take the continuum time limit
this is sufficient. If $C_i \neq C_{i+1}$ then the two configurations
can only differ in the occupation numbers of either the spin up or the 
spin down impurity fermion and the corresponding spin of one of the band
fermion levels with momentum say ${\bf k}$. Further both spins cannot hop 
simultaneously! Thus, if these constraints are met then
\begin{equation}
<C_i|{\rm e}^{-\tau H}|C_{i+1}> = \tau |V_{\bf k}|,
\end{equation}
otherwise the matrix element is zero and is disallowed. Our directed loop 
update begins by choosing a point, with $50\%$
probability on the impurity and $50\%$ probability on the other
sites chosen randomly on one of the spin layers. Then with probability
half the path enters the vertex either in the positive time direction
or the negative time direction. Using a set of rules that governs 
the exit of the update path at each ``vertex'' given the entrance of 
the path, the loop grows until it finds the starting point, where the
update ends. The occupation numbers are changed while the loop is being
constructed. 

\begin{figure}
\centerline{\psfig{figure=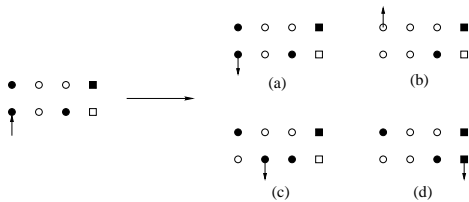,height=0.3\hsize}}
\caption{When the directed loop enters a vertex (shown on the left)
one can produce many exit paths (some of them are shown on the
right after the flip in the occupation numbers).
Circles indicate electron levels in the band and squares 
represent the impurity (filled symbols indicate occupied sites). 
Each figure only shows a few relevant electron sites. For the Anderson 
model discussed in the text (c) and (d) are forbidden to order $\tau$. 
Both (a) and (b) are allowed but we can choose the weight of (a) to be 
zero while satisfying detailed balance.} 
\label{fig:vertex0} \end{figure}

\begin{figure}
\centerline{\psfig{figure=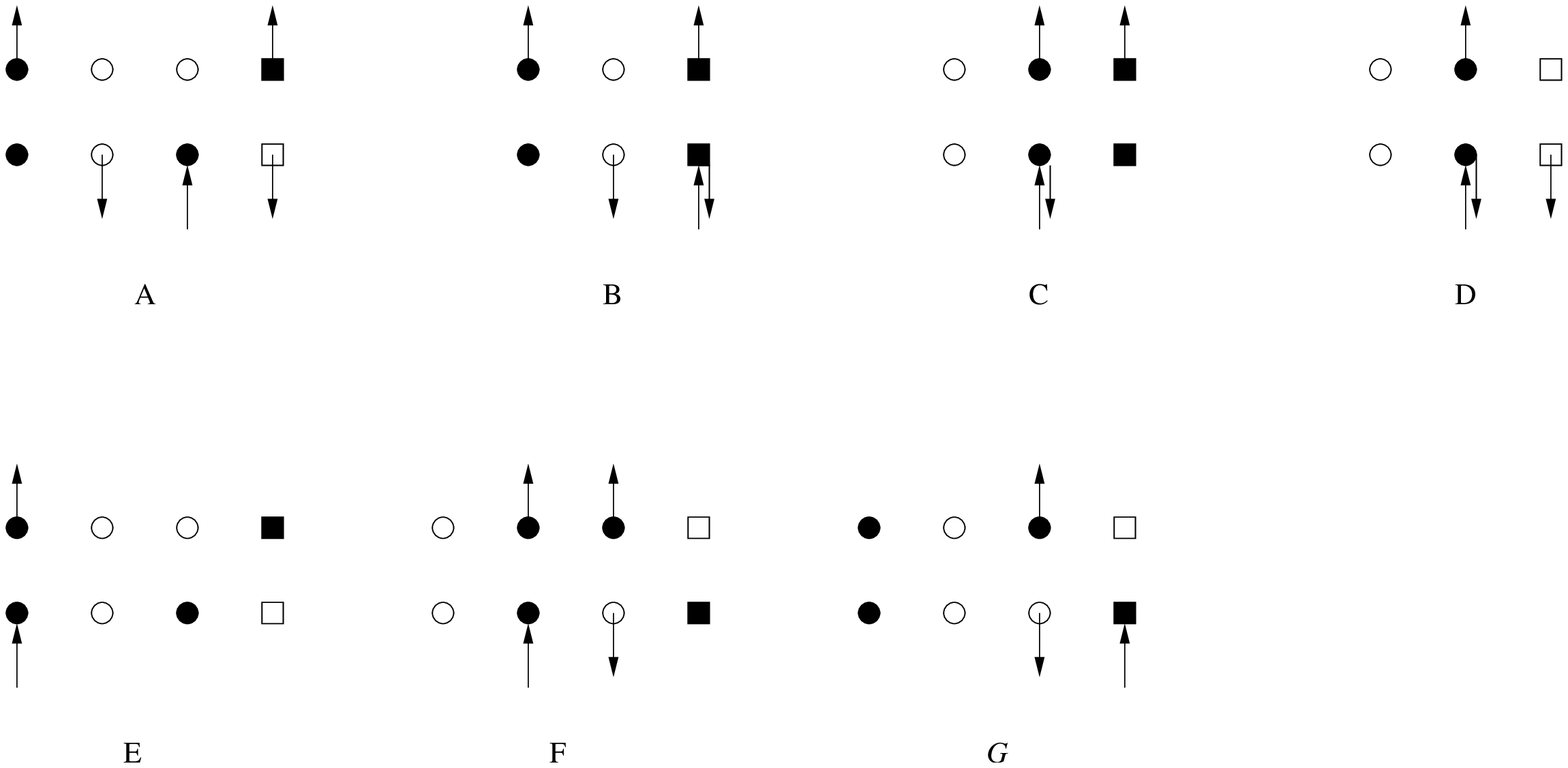,height=0.4\hsize}}
\caption{Assignments of directed-loop segments. All possible vertices 
in which the entrance site is occupied are shown. Vertex E was considered 
in Fig.\ \ref{fig:vertex0}. However, unlike Fig.\ \ref{fig:vertex0} all
the exit paths are shown together without the flip in occupation numbers.} 
\label{fig:vertexall} \end{figure}

Since the Hamiltonian commutes with the total spin operator and the total
fermion number operator, the rules that generate the loop must satisfy the
appropriate conservation laws. In addition, since we will ultimately
take the $\tau\rightarrow0$ limit, processes of order $\tau^2$ should not
be considered. Thus, only one hopping can occur at a vertex. Using these 
constraints one can determine all the allowed processes in a given
vertex. This is illustrated in Fig.\ \ref{fig:vertex0} in which the directed
loop enters the vertex through a filled electron state in the band
with a definite spin. On the right, four occupation number conserving exit 
paths ((a),(b),(c),(d)) are shown. It is easy to see that, up to order 
$\tau$, only (a) and (b) are allowed. Once all the possible paths are
determined, all the processes are given probability weights such that
they satisfy detailed balance. For example in this case the bounce weight 
(a) can be chosen to be zero, so that the continuation process (b) occurs
with probability one. For vertices where the path enters an occupied site, all 
possible loop segment assignments are shown in Fig.\ \ref{fig:vertexall}. The
example considered in Fig.\ \ref{fig:vertex0} is shown as vertex E in 
Fig.\ \ref{fig:vertexall}. Let us now discuss the other vertices.

First, consider a vertex where a hopping occurs at site $k$ and the loop 
enters the vertex at site $k$ (vertex A in Fig.\ \ref{fig:vertexall}).
There are a total of $N+1$ different possible exits in this vertex 
where $N$ is the total number of band energy levels. Let $A_{kq}$ be the 
weight for the path to exit at another band electron level $q$. We choose 
\begin{equation}
A_{kq}=\tau\, \frac{{\rm min}[|V_k|,|V_q|]}{N+1}.
\end{equation}
The probability for this process will then be given by 
$P_{kq} = A_{kq}/(\tau |V_{\bf k}|)$. Since $A_{kq}$ is symmetric 
in $k$ and $q$ detailed balance is satisfied. The remaining probability
must be the probability for the loop to exit at the impurity. Since there 
are two possible paths, for simplicity the weight for each of these processes
is chosen to be
\begin{eqnarray}
A_{kd}=\frac{1}{2}(\tau |V_k|-\sum_{q\neq  k} A_{qk}).
\end{eqnarray}
Then, the probability for these process is $A_{kd}/\tau |V_{\bf k}|$.
The factor $N+1$ guarantees that all of $A_{qk},A_{dk}$ are positive 
numbers. Our choice of weights is such that in the case where all 
$|V_{\bf k}|$ are the same all possible processes are equally likely.
Note that the loop does not bounce back at this vertex,
and there is a unique direction for the loop at each of the band levels.

In the case of vertex B the incoming path is on the impurity. The outgoing 
path can be at one of the electron band levels with momentum $k$ which has 
the weight $A_{dk}$ which will be chosen to be equal to $A_{kd}$ to satisfy 
detailed balance. In this case there is a possibility for the
loop to continue forward or to bounce back. In order to fix the weights of 
these processes one compares the weight of the original vertex with that 
of the vertex obtained if the loop continues to go forward. 
If the forward continuation produces a vertex of the smaller weight, 
the bounce weight is chosen to be the absolute value of difference of the 
two weights; otherwise the bounce weight is zero. There are two cases to be 
considered: for the case in which the impurity site contains an electron 
with the opposite spin (to the spin at which the path is being constructed), 
the bounce weight is $|\tau (\epsilon_d+U)|$, and in the other case 
$|\tau (\epsilon_d)|$. This prescription along with the normalization 
condition fixes the weight of the continuation process. 

In vertices C, D and G every weight for the exit path hopping to or out of 
the impurity is chosen as $|\tau V_k/2|$ so that the forward and backward 
probabilities in G are equal. Continuation and bounce weights in vertices C 
and D
are determined in the same way as for vertex B; the bounce weight turns out
to be $|\tau \epsilon_{\bf k}|$ if forward continuation lowers the weights,
otherwise zero. In G, there is no bounce back. In vertex F 
a hopping from site $k$ to site $q$ occurs with weight 
$\tau\frac{\min[|V_k|,|V_q|]}{2}$. Finally, as discussed earlier, in 
vertex E the path is always forced to continue. 

The above rules can easily be extended to the case in which the directed
loop enters a vertex on an empty site. Although the above rules satisfy
detailed balance, they are in no way unique. We chose the above rules
after testing a few other possibilities since they were similar in efficiency
if not better than other ones. It has been suggested that if the bounce 
probability is large then the algorithm is likely to become inefficient, 
since the proposed change is being rejected. In our case a large 
bounce weight in electron sites far away from the Fermi surface is natural 
since the electron sites there either remain occupied or empty most 
of the time. On the other hand, a large $\epsilon_d$ or $U$ 
leads to a large bounce probability in the vertex B at the impurity. Thus, 
a large $\epsilon_d$ or $U$ may cause inefficiencies in our algorithm. One can 
choose a different set of weights to reduce the bounce weights by taking 
into account $\epsilon_k$ at all values of $k$, but determining the weights
in this case becomes difficult. We have tested a few different set of weights 
which gives less bounce back at the impurity and band electrons. Unfortunately, our attempts 
have not improved the efficiency of the algorithm further. So here we report 
on the results using simplest algorithm discussed above.

Until now the limit of $\tau\rightarrow 0$ was not taken. As $\tau$ is 
taken to the zero, one gets the continuous time version. From 
$A_{qk}$, $A_{dk}$, and the bounce weights, one can easily evaluate 
the decay rates for the continuous time simulation. The Monte Carlo 
simulation in continuous time proceeds as follows: First, we pick a 
starting time, spin, and path direction. For the starting site, the 
impurity site is picked more often. Typically 50\% of the starting points
are at the impurity and the remaining $50\%$ on the levels in the band 
with equal probability \cite{expdis}.
The path for the loop continues in time 
until a decay occurs into one of the possible vertices. Then, the new 
level and the direction are determined by the exit process. 
If a path hits a time slice where the configuration changes before
a decay occurs, then the vertex at that time slice is used to
decide the exit process. The loop update continues until it closes.
As the loop is constructed the occupation states along the loop are flipped.

\section{The Multilevel Algorithm}

In a given configuration $C$ electrons hop between the band and the 
impurity site so that in a periodic configuration in imaginary time, 
the electrons permute their positions. Due to the Pauli principle, 
this causes configurations to have a positive or a negative sign. 
This is the reason for the factor $\sigma[C]$ in Eq.(\ref{boltz}). 
Any physical quantity $O$ can be computed using
\begin{equation}
\langle O \rangle = \frac{1}{Z}{\rm Tr}\,O e^{-\beta H} = 
\frac{\sum_C\,O\, \sigma[C] W(C)}{\sum_C\, \sigma[C]\, W(C)} =
\frac{\langle O \sigma\rangle}{\langle \sigma\rangle},
\end{equation}
where the final expectation values are computed using the Monte Carlo 
algorithm discussed in the previous section that generates configurations
with probability $P(C)$ defined in Eq.(\ref{prob}). Unfortunately, as the
temperature decreases, both the numerator and the denominator decrease 
exponentially which makes the calculations of fermionic observables at 
low temperatures extremely difficult. Thus one needs an efficient method
to compute exponentially small numbers by averaging large positive
and negative numbers, a problem that is generically referred to as 
the {\em Sign Problem}.

Recently, a clever trick referred to as the multi-level algorithm was 
discovered in the context of lattice QCD to compute exponentially small 
numbers \cite{Lue}. In particular it was possible to compute the
potential $V(R)$ between quarks and anti-quarks, by computing the 
corresponding exponentially small Boltzmann weight $\exp(-V(R)/T)$ at a 
temperature $T$. In lattice QCD this quantity can be computed by averaging
the Wilson loop which is typically of order $1$ for a given configuration. 
It was shown that the multi-level algorithm could compute averages of
Wilson loops that were as small as $10^{-20}$. The basic idea was to write 
the observable, in this case the Wilson loop, as a product of many terms 
such that each of the terms is not very small even though the product is 
very small. In this article we show that a similar approach can be applied 
to compute the average sign in the fermionic problem using the directed-loop 
algorithm.

In order to apply the multi-level algorithm let us divide the Euclidean
time $\beta$ of the lattice into $2^{K-1}$ parts with the same time 
thickness. Let us denote the sub-lattice configuration of fermions inside 
each of these parts by $C_{s_i}, i = 1,2,...,2^{K-1}$ and the boundaries 
between the sub-lattices by $C_{b_\tau}$ where 
$\tau=0,\varepsilon,2\varepsilon,...,\beta$ 
represents the Euclidean times at the boundaries. Periodic boundary conditions
in time means that $C_{b_0} = C_{b_\beta}$. Now, the boundary configurations 
$C_b$ and sub-lattice configurations $C_s$ determine the entire configuration
$C$. The probability $P(C)$ can then be expressed as
\begin{equation}
P(C) = P(C_b) \prod_{i=1}^{2^{N-1}} P(C_b,C_{s_i}),
\end{equation}
where $P(C_b)$ is the probability of finding the configuration $C_b$ on the
boundaries and $P(C_b,C_{s_i})$ is the conditional probability of finding the
configuration $C_{s_i}$ given the boundary configurations $C_b$. Clearly,
$P(C_b,C_{s_i})$ depends only on the boundaries that bound the 
$i$th sub-lattice. Since the sign of a configuration $\sigma[C]$ can be 
written as a product of sign factors coming from each of the 
sub-lattices, the average sign can be written as
\begin{equation}
\langle \sigma \rangle = \sum_{C_b,C_{si}}\,P(C_b) \prod_i\, 
\sigma_i\,P(C_b,C_{si}),
\end{equation}
where $\sigma_i$ is the sign that comes from the sub-lattice $i$. Now 
$\sum_{C_{si}} \sigma_i\,P(C_b,C_{si})$ is just the average sign of the 
$i$-th sub-lattice with a fixed boundary configuration. So
\begin{equation}
\langle \sigma \rangle = \sum_{C_b}\,P(C_b) 
\prod_i\, \langle \sigma_i \rangle({C_b}),
\end{equation}
where $\langle \sigma_i \rangle(C_b)$ is the average sign of the 
sub-lattice $i$ with the boundary $C_b$. 

The multi-level algorithm proceeds as follows:
First, to generate a sequence of boundaries $C_b$, one updates the entire 
lattice. Then, with a fixed boundary configuration $C_b$, one generates 
a subsequence of $N_s$ configurations for each sub-lattice. The
directed loop algorithm is well suited for this update; when the directed-loop
encounters the fixed time slice the path is forced to bounce back! Clearly 
one can estimate the average sign $\langle \sigma_i \rangle$ of each 
sub-lattice 
independently using the $N_s$ configurations and then use their product to 
compute $\langle \sigma \rangle$. It should be emphasized that although the
sub-lattice averages are not exact there is no systemic errors in this
approach. $N_s$ is determined empirically so as to make the calculation
efficient. The sub-averages do not have to be calculated more accurately than 
the size of the fluctuations due to change in the boundaries. Once the
sub-lattices are updated the entire configuration is again updated to generate 
a new set of boundary configurations $C_b$. Repeating this process, a series 
of sign measurements are generated. The final result of the sign is obtained 
by averaging these measurements. The statistical noise in the sign is reduced 
because effectively one is summing over $N_s^{2^{K-1}}$ configurations for 
each of these measurements.

\begin{figure}
\centerline{\psfig{figure=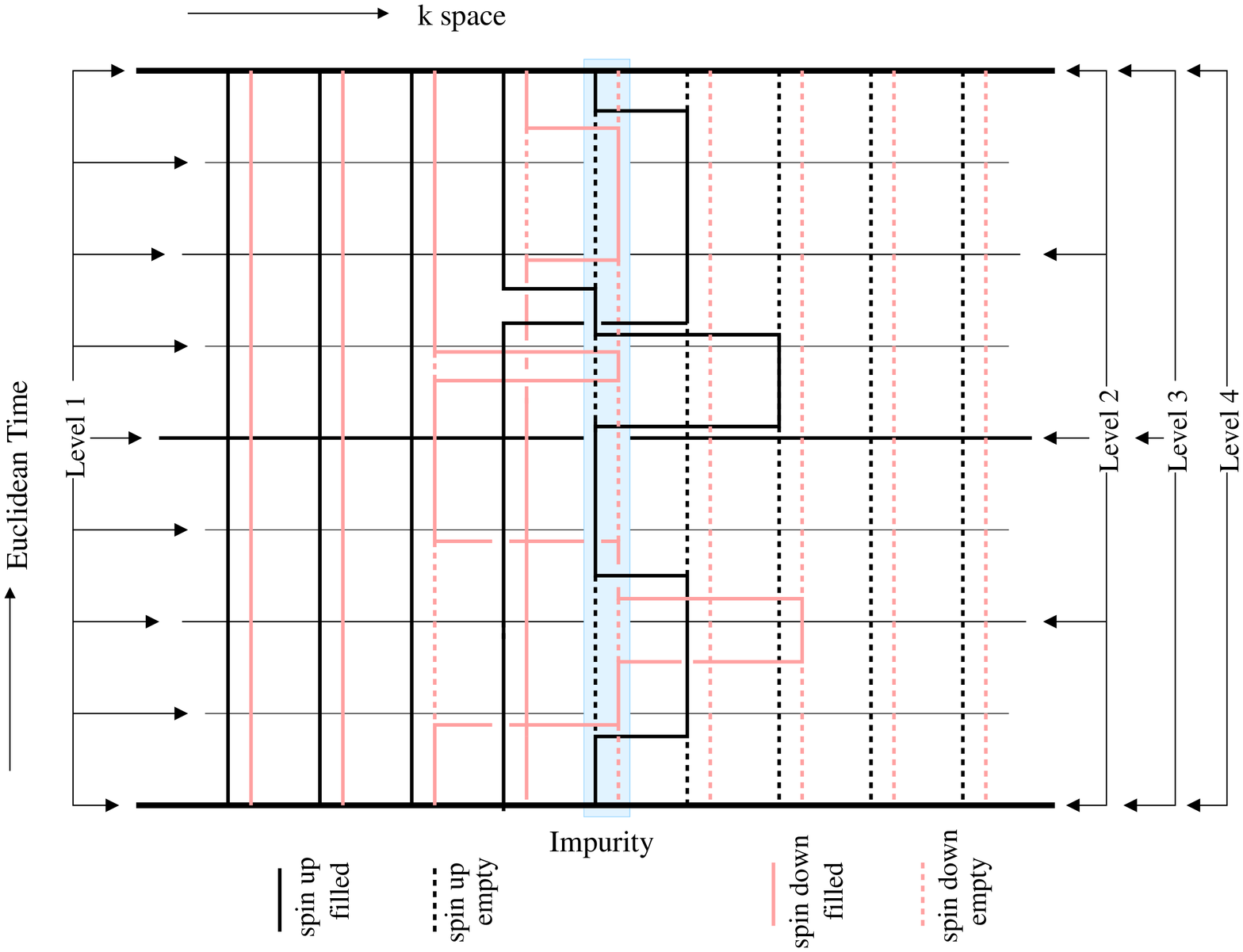,width=0.8\hsize}}
\caption{A schematic description of the four level algorithm.} 
\label{fig:multilevel} \end{figure}

In the multi-level algorithm one can also build nested levels of update. 
In particular one can performs a $K$ level update if the Euclidean time 
direction is divided into $2^{K-1}$ parts. During the update at the first 
level all time slices $t=0,\varepsilon,...(2^{K-1}-1)\varepsilon$ are held 
fixed, in the update as discussed above. At this level $N_s$ loop updates
are performed between the fixed time-slices. At the second level only time 
slices at $t=0,2\varepsilon,4\varepsilon,...$ are held fixed. After every 
update at the second level, the full first level update is performed. This
procedure is repeated $N_s$ times. This process is repeated at higher levels;
such that at each highest level update, all the complete lower level updates
are performed. Thus, at the $K$th level the whole lattice is updated with 
just the $t=0$ time-slice held fixed. One can extend the ideas of measuring 
the sign for a single level update as discussed above, to the $K$-level 
update. After summing over the signs of all the configurations generated 
during the $K$ level update one computes a value of the sign. If at each 
level one performs $N_s$ updates for each sub-lattice, one performs on the 
order of $(N_s)^K \times 2^{K-1}$ loop updates after the $K$th-level update. 
Since $\beta=2^{K-1} \varepsilon$, then for a fixed $\varepsilon$ if the effort
for a single loop update remains fixed, the effort to compute the sign
for a full $K$-level update grows as a power of $\beta$. As more levels 
are introduced,  it takes longer for the measurement of sign. In Fig.\  
\ref{fig:multilevel} we show a schematic description of the multi-level idea. 

One can apply the multi-level technique to other observables which can
be written as product of quantities on each of the sub-lattices. We call
such observables as being compatible with the multi-level algorithm. The 
optimum number of levels should be determined empirically. This number 
can depend on the observable to be calculated. We also found that the 
full multi-level algorithm is the most efficient for fermion sign problems
due to the large oscillations. 

\section{Observables}

Since the Monte Carlo update is performed in the occupation number basis, all
diagonal observables $O[n]$ that are functions of the occupation numbers 
can easily be calculated using the formula
\begin{equation}
\langle O \rangle = 
\frac{\langle O[n] \sigma[C]\rangle}{\langle \sigma[C] \rangle}.
\end{equation}
Average occupation number of a level is one of the observables which belongs 
to this class. Another important diagonal observable in the Anderson impurity 
model is the local susceptibility 
\begin{eqnarray}
\chi_{\rm lo} = \int_0^\beta \langle s(\tau)s(0) \rangle,
\end{eqnarray}
which can be obtained from configurations of the impurity as a function of 
imaginary time. This method of computation can only be reliable for regions of 
temperature where the sign problem is mild or when the multi-level algorithm 
discussed above is applicable and useful. Both, the average occupation numbers 
and the local susceptibility, are compatible with the multi-level algorithm 
and hence the algorithm can be used to alleviate the sign problem in 
calculating these quantities.

The quantity more directly relevant to experiment is the impurity 
susceptibility, $\chi_{\mathrm{im}}$, which is the total susceptibility 
minus the free 
susceptibility:
\begin{eqnarray*}
\chi_{\mathrm{im}} = \chi_{tot} - \sum_i \chi_i,
\end{eqnarray*}
where $\chi_{i}$ is the susceptibility from the $i$-th free electron site.
In the determinantal Monte Carlo (the Hirsch and Fye algorithm) the impurity 
susceptibility is very difficult to calculate due to statistical noises 
in the simulation \cite{Fye}. The problem is that one has to calculate the total 
susceptibility for the Anderson Hamiltonian and then subtract the 
susceptibility for the free case from it. The total susceptibility is a 
quantity of order $N$ (the number of band electron sites), but the impurity 
susceptibility is of order 1. So one has to calculate the total susceptibility 
with the error of order 1 or less, which is extremely difficult for large 
lattice $N$. From the Clogston-Anderson compensation theorem \cite{And}, for a large bandwidth with 
a flat energy density and equal hopping amplitudes, one can expect that the 
local susceptibility is equal to the impurity susceptibility. But in the 
study of mesoscopic fluctuations the two susceptibilities can be quite 
different \cite{San}. One of the main advantages of our method is that
with our algorithm the impurity susceptibility can be measured with 
significantly reduced statistical noise. Let us now discuss how we can compute
$\chi_{\mathrm{im}}$.

For configurations in the occupation number basis generated during the update,
the hopping of electrons occurs at only a small number of electron sites.
The rest of the energy sites appear to be free; an advantage of working in
the ``momentum'' space lattice. Suppose that  $N_{\rm hop}$ sites are involved 
in the hopping for a given configuration $C$. Denote the configuration of 
those sites by $C_{\rm hop}$, and the free part by $C_f$. Then, one can express 
the probability of $C$ as $P(C)=P(C_{\rm hop})P(C_{\rm hop},C_f)$, where 
$P(C_{\rm hop},C_f)$ is the probability of $C_f$ with a given $C_{\rm hop}$.
The impurity susceptibility can be expressed with $P(C_f)$ and 
$P(C_{\rm hop},C_f)$ as
\begin{eqnarray*}
\chi_{\mathrm{im}} &=& \frac{1}{\langle \sigma\rangle} 
\sum_{C_{\rm hop}}\, P(C_{\rm hop}) \sigma[C_{\rm hop}] \\
&\times&
\{\chi(C_{\rm hop})\\ 
 & &\, + \sum_{C_f} P(C_{\rm hop},C_f)[\chi (C_f)
- \sum_{i\in C_f} \chi_i - \sum_{i\in C_{\rm hop}} \chi_i] \}, 
\end{eqnarray*}
where $\chi (C_{\rm hop})$, $\chi (C_f)$ are the susceptibilities from
sites that contain electron hops and those that appear free in a 
given configuration. Since $C_f$ contains no hop, one can see that
\begin{equation*}
\sum_{C_f} P(C_{\rm hop},C_f)[\chi (C_f)- \sum_{i\in C_f} \chi_i] = 0.
\end{equation*}
Using this one can find that the impurity susceptibility is given by
\begin{eqnarray}
\chi_{\mathrm{im}} = \frac{1}{\langle \sigma\rangle}
\ \ \Bigg\langle \sigma[C_{\rm hop}][\chi(C_{\rm hop}) - 
\sum_{i\in C_{\rm hop}} \chi_i]\Bigg\rangle,
\label{impsus}
\end{eqnarray}
where the free susceptibility of the sites in $C_{\rm hop}$ only needs to be 
subtracted. In this method the size of statistical fluctuation of 
the measurements of the impurity susceptibility is of order of the number of 
electron sites in which the hopping occurs for a configuration. So the 
statistical noise in this method are substantially reduced. The observables
that go into Eq.(\ref{impsus}) are again compatible with the multi-level
algorithm.

Unfortunately, we have found that the above technique is still noisy in
practice. Interestingly, one can reduce the statistical noise in 
$\chi(C_{\rm hop})$ further by using the technique of improved estimator that 
is commonly used in cluster algorithms. In this method one identifies all 
spin clusters for a given configuration $C_{\rm hop}$ that can be flipped 
independently and performs a partial average over these cluster flips.
Unfortunately, this step is not compatible with the multi-level algorithm. 
But for the case where the sign problem is moderate the impurity 
susceptibility can be computed very efficiently using our algorithm.

\section{Results}

In this section we discuss results from the simulation of the Anderson 
Hamiltonian given in Eq.(\ref{aham}) with the algorithm described in 
the previous sections. First, let us focus on the calculation of the 
average sign using the multi-level 
algorithm. For this purpose we choose $N=750$ equally spaced energy levels 
with a bandwidth of $2D=10$. We choose $V_{\bf k} = V$ such that 
$\Gamma= \pi \rho V^2 = 0.5$ and $U=2$. We have studied three different temperatures 
by choosing $\beta$ to be 40, 80 and 160 in order to see the effectiveness of 
the multi-level algorithm in the computation of the average sign. The 
sub-lattice thickness $\varepsilon$ is chosen to be $10$ so that at 
$\beta=40$ we have four, at $\beta=80$ we have eight and at $\beta=160$ 
contains sixteen sub-lattices.

For the sub-lattice of thickness $\varepsilon = 10$, we found that 
$N_s = 10$ updates was necessary to get an reasonable estimate of
the average sign of the sub-lattice. To complete one full cycle of the 
all multi-level updates, $(2N_s)^K/2$ loop updates are required where 
$K=3$ for $\beta=40$, $K=4$ at $\beta=80$ and $K=5$ at $\beta=160$.
In addition, at each higher level sub-lattices were updated 4 times to 
generate new boundaries between sub-lattices.

\begin{figure}
\centerline{\psfig{figure=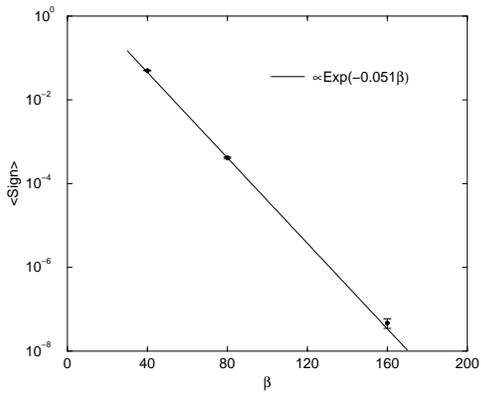,height=0.6\hsize}}
\caption{Average signs versus $\beta$ for
$N=750$, $D=5$, $U=2$, $\epsilon_d=-1$ and $\Gamma=0.5$.} \label{fig:ave_sign}
\end{figure}

At $\beta=40$ and 80, we computed $\langle \sigma\rangle$ to be
$4.99(11)\times 10^{-2}$ and $4.13(16)\times 10^{-4}$, respectively, where the
errors are of the order of a few percent.
At $\beta=160$ the sign average is so small and the projected time 
is so long that the simulation was stopped when the error was about 
30 percent. The average sign at $\beta=160$ was $4.7(1.2)\times 10^{-8}$.
The computational time taken for these results are 3hrs., 92hrs., and 
2000hrs., respectively. For the $\beta=160$, 6 CPUs were used with 
different random number seeds to collect the data. In Fig.\ 
\ref{fig:ave_sign} we plot the average sign as a function of $\beta$ and 
we see that all the values fall nicely on an exponential form
as expected.

\begin{figure}
\centerline{\psfig{figure=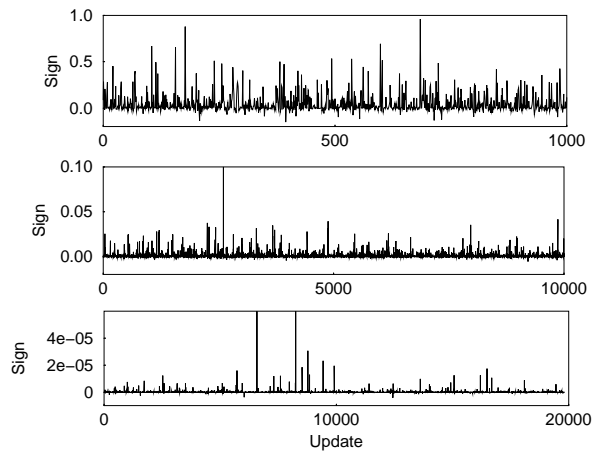,height=0.7\hsize}}
\caption{Monte Carlo sequence of fermion signs for $\beta=40$, 80 and 160 from 
the top.} \label{fig:signs}
\end{figure}

\begin{figure}
\centerline{\psfig{figure=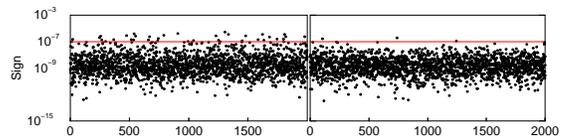,height=0.2\hsize}}
\caption{The first 2000 of the positive(left) and negative signs in 
logarithmic scale of the $\beta=160$ run.} \label{fig:signs_pos_neg}
\end{figure}

Now the biggest question to answer is whether the multi-level algorithm is
useful. We would first like to point out that it is still difficult to
compute the average sign with reasonable errors at very small temperatures. 
The required effort grows as at least a large power of $\beta$, and we cannot
rule out an exponential growth at the moment. However, the fact that 
for $\beta=160$ we could compute numbers of the order of $10^{-8}$ itself is
an indication that some progress has been achieved. Without the multi-level
algorithm this would have been impossible. If we look at the
individual values of the sign computed by the multi-level algorithm after
each update we learn something further. Fig.\ \ref{fig:signs} shows these 
values of the signs for different values of $\beta$. We see that using the 
multi-level algorithm the values of the signs are dominated by positive values. 
For example one can see that at $\beta=160$ the average sign has most 
contributions from the positive side. We show this in Fig.\ 
\ref{fig:signs_pos_neg} by focusing on the first 2000 positive and negative
values. We see that although the very small values come with equal weight
between positive and negative values, there are a few positive numbers that
dominate over the negative numbers. The multi-level algorithm has allowed 
us to find configurations whose signs, when summed up, leads to a very small
number which can either be positive or negative but occasionally it also
leads to numbers which are orders of magnitude larger but mostly positive.
The large error in the average sign is mainly due to these large fluctuations
but always in the positive direction. In the next section we will discuss
why this observation is interesting.

\begin{figure}
\centerline{\psfig{figure=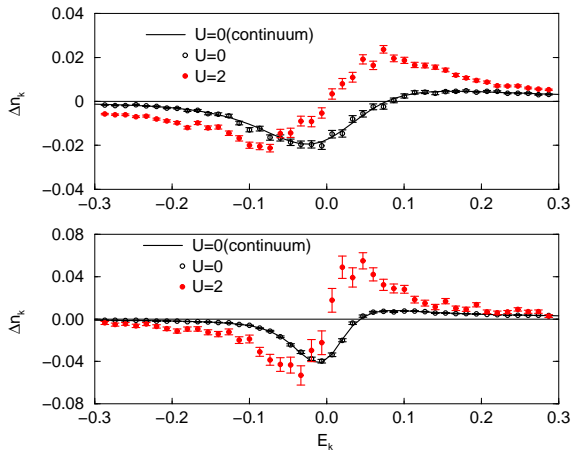,height=0.7\hsize}}
\caption{Occupation numbers minus the free Fermi function at $\beta=25$(top) 
and 50(bottom) for $N=750$, $D=5$, $\epsilon_d=-1$, and $\Gamma=0.5$.
\label{fig:occup}}
\end{figure}

In order to show that the new algorithm is indeed interesting, for the 
parameters chosen above we compute the average occupation numbers for 
the various energy levels. In Fig.\ \ref{fig:occup}
we plot the differences between the average occupation 
numbers in the interacting cases and the free cases at the temperatures 
$\beta=25$ and $50$. The open circles give the results in the cases where 
$U=0$ in which case one can obtain the result also from the exact
Green functions assuming the density of energy levels is smooth
(solid line). Thus, we see that our algorithm can indeed produce results 
in good agreement. For the case of $U=2$, the Kondo temperature for this 
system is roughly $T_k = 0.09$ as seen from the numerical renormalization 
method \cite{Wil,Hir}. In this case $\beta=25$ and $50$ correspond to a 
temperature of about $T_k/2$ and $T_k/4$. We see that indeed
the Kondo resonance appears as expected.

\begin{figure}[b]
\vskip0.4in
\centerline{\psfig{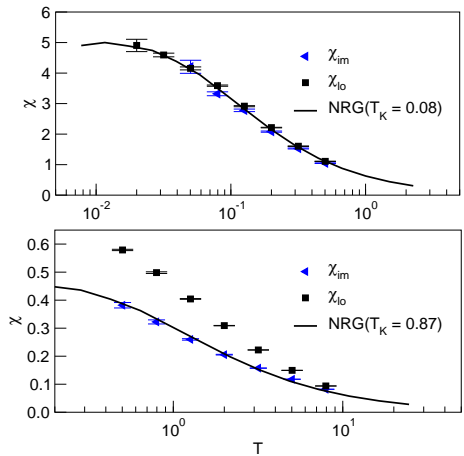}}
\caption{Local and impurity susceptibilities for $U=2$(top) and 
$U=25$(bottom) with $U/\Gamma=4$ fixed.} \label{fig:chi}
\end{figure}

Finally, we focus on the local and impurity susceptibilities. To check 
the Clogston-Anderson compensation theorem, a larger bandwidth $2D=20$ is 
chosen with $N=2000$. For $U=2(\epsilon_d=-1)$,  $\chi_{\mathrm{lo}}$ and 
$\chi_{\mathrm{im}}$ are shown 
in Fig.\ \ref{fig:chi}. For the local susceptibility, with the multi-level 
method we were able to calculate the local susceptibility at a much lower 
temperature than the impurity susceptibility. We find that 
$\chi_{\mathrm{lo}}$ 
and $\chi_{\mathrm{im}}$ are in reasonable agreement as expected from the 
Clogston-Anderson compensation theorem. We also compare our result with
the NRG curve (solid line) obtained for $T_k=0.08$ which passes through most 
of the data points.

Since our simulations are in continuous Euclidean time, we can simulate 
a large $U$ without increasing the discretization error. In the limit of 
large $U$, the Anderson model converges to the Kondo model. In the Kondo 
model, band electrons and impurity spin interact with the coupling $J$. 
From the Schrieffer-Wolff transformation \cite{SW}, the effective coupling 
$J$ of the Anderson Hamiltonian for a large $U$ is 
$\frac{8\Gamma}{\pi U\rho}$.
In order to go towards the Kondo limit we fix $U/\Gamma=4$ and 
study the case where $U=25(\epsilon_d=-12.5)$. 
The local and impurity susceptibilities are plotted in Fig.\ \ref{fig:chi}.
We see that now these two are completely different. We attribute this
difference to the fact that $U\gg D$ in which case the compensation 
theorem is no longer valid.

We have also computed the various observables discussed in this article
for the Hamiltonian that contains mesoscopic fluctuations. We find that
typically we can use our new method to compute quantities for temperatures
as low as $T\sim T_k/4$. Using the multi-level technique we can also go
down to temperatures of about $T\sim T_k/10$.

\section{Conclusion}

In this article we have investigated a new algorithm for a model involving
a band of fermions interacting with a single impurity in the occupation 
number basis in ``momentum'' space. We use the efficient directed loop 
algorithm to update the configurations and absorb the sign into observables. 
We find that the sign problem is mild down to temperatures of order  
$T_k/4$. Further, the new approach allows us to explore a new multi-level 
algorithm to compute average signs efficiently. We were able to compute signs 
of the order of $10^{-8}$ with moderate effort. The new approach also
allows us to calculate certain quantities like the impurity susceptibility 
more efficiently than conventional Monte Carlo methods. Finally, our 
algorithm can easily be extended to several impurities.

The average of the sign over configurations that are generated in the 
multi-level algorithm fluctuates between small values which can be both 
positive and negative and large values which are orders of magnitude 
larger but always positive. The effort for this grows as $(2N_s)^K/2$ 
where $\beta = 2^{K-1}$. Although this does not solve the
sign problem completely, since the positive numbers can still fluctuate a 
lot, perhaps half of the sign problem has been solved. If this is true then
we think this is an exciting step in the solution to the full sign problem 
based on the recent progress in solving certain sign problems using the 
meron cluster algorithm \cite{Cha99}. There it was possible to rewrite the 
partition function in terms of configurations where the Boltzmann weight 
was either zero or positive. Thus all negative signs were eliminated. The 
second step was algorithmic when all zero configurations were eliminated 
in an accept reject step. An intriguing question is whether something 
similar can be achieved in the present case. We leave this question for
future research.

\section{Acknowledgments}
We thank U.J.~Wiese and R.K.~Kaul for useful discussions.
This work was supported in part by the NSF grant 
(DMR-0103003).


\begin{thebibliography}{99}
\bibitem{Bla81} R.~Blankenbecler, D.J.~Scalapino and R.L.~Sugar,
Phys. Rev. D24, 2278 (1981).
\bibitem{Hir} J.E.~Hirsch and R.M.~Fye, 
Phys. Rev. Lett. 56, 2521 (1986).
\bibitem{Wie93} U.J.~Wiese, 
Phys. Lett. B311, 235 (1993).
\bibitem{Kaw94} N.~Kawashima, J.~Gubernatis and H.G.~Evertz,
Phys. Rev. B50, 136 (1994).
\bibitem{Cha99} S.~Chandrasekharan and U.J.~Wiese, 
Phys. Rev. Lett. 83, 3116 (1999).
\bibitem{Cha02} S.~Chandrasekharan and J.C.~Osborn, 
Phys. Rev. B66, 045113 (2002).
\bibitem{CMclassics}
J.M. Ziman, {\it Principles of the Theory of Solids} (Cambridge
University Press, Cambridge, 1972) chapter 10; P. Phillips, {\it
Advanced Solid State Physics} (Westview Press, Cambridge MA, 2003)
chapters 6 and 7.
\bibitem{qdots}
L Kouwenhaven and L. Glazman, Physics World, January 2001, pp. 33-38;
M. Pustilnik, L.I. Glazman, D H. Cobden, and L.P. Kouwenhoven, Lecture
Notes in Physics {\bf 3}, 579 (2001),
http://www.arxiv.org/abs/cond-mat/0010336;
L. Borda, G. Zarand, W. Hofstetter, B.I. Halperin, and J. von Delft,
Phys. Rev. Lett. {\bf 90}, 026602 (2003).
\bibitem{ribhu}
R.K. Kaul, D. Ullmo, S. Chandrasekharan, and H.U. Baranger,
preprint, cond-mat/?? (2004).
\bibitem{dmft}
G. Kotliar and D. Vollhardt, Physics Today {\bf 57}, no. 3, pp. 53-63
(March 2004);
A. Georges, G. Kotliar, W. Krauth, and M.J. Rozenberg
Rev. Mod. Phys. {\bf 68}, 13 (1996).
\bibitem{And} P.W.~Anderson, 
Phys. Rev. 124, 41 (1961).
\bibitem{Hew} A.C. Hewson, {\it The Kondo Problem to Heavy Fermions} 
(Cambridge University Press, Cambridge, 1993).
\bibitem{Wig} P.B.~Wiegmann, 
Phys. Lett. A31, 163 (1981). 
\bibitem{Kaw} N.~Kawakami and A.~Okiji, 
Phys. Lett. A86, 483 (1981). 
\bibitem{Wil} H.R.~Krishna-murthy, J.W.~Wilkins and K.G.~Wilson, 
Phys. Rev. B21, 1003 (1980).
\bibitem{Sand} O.F.~Syljuasen and A.W.~Sandvik, 
Phys. Rev. E46, 046701 (2002).
\bibitem{Bea96} B.~Bernard and U.J.~Wiese,
Phys. Rev. Lett. 77, 5130 (1996).
\bibitem{expdis} One can use other distributions: e.g.,
a exponentially decaying function away from the Fermi level. 
\bibitem{Lue} M.~L\"{u}scher and P.~Weisz, 
JHEP 0109, 010 (2001).
\bibitem{Mak98} C. H.~Mak, R.~Egger, and H.~Weber-Gottschick
Phys. Rev. Lett. 81, 4533 (1998).
\bibitem{Fye} R.M.~Fye and J.E.~Hirsch, Phys. Rev. B38, 433 (1988).
\bibitem{San} G.E.~Santoro and G.F.~Giuliani, 
Phys. Rev. B44, 2209 (1991).
\bibitem{SW} J.R Schrieffer and P.A. Wolff, Phys. Rev. 149, 419 (1966).
\end{thebibliography}
\end{document}